\documentclass[12pt]{spieman}  
\usepackage{amsmath,amsfonts,amssymb}
\usepackage{graphicx}
\usepackage{setspace}
\usepackage{tocloft}
\usepackage{subcaption}
\usepackage[numbers]{natbib}

\usepackage[symbol]{footmisc}

\title{Development and characterization of a fast and low noise readout for the next generation X-ray CCDs}

\author[a]{Tanmoy Chattopadhyay}
\author[a]{Sven Herrmann}
\author[a]{Peter Orel}
\author[a]{R. Glenn Morris}
\author[b]{Gregory Prigozhin}
\author[b]{Andrew Malonis}
\author[b]{Richard Foster}
\author[c]{David Craig}
\author[c]{Barry E. Burke\footnote[2]{Deceased}}
\author[a,d,e]{Steven W. Allen}
\author[b]{Marshall Bautz}

\affil[a]{Kavli Institute of Astrophysics and Cosmology, Stanford University, 452 Lomita Mall, Stanford, CA 94305, USA}
\affil[b]{Kavli Institute for Astrophysics and Space Research, Massachusetts Institute of Technology, Cambridge, MA USA}
\affil[c]{MIT Lincoln Laboratory, Lexington, MA, USA}
\affil[d]{SLAC National Accelerator Laboratory, 2575 Sand Hill Road, Menlo Park, CA 94025, USA}
\affil[e]{Department of Physics, Stanford University, 382 Via Pueblo Mall, Stanford CA 94305, USA}

\cftpagenumbersoff{figure}
\cftpagenumbersoff{table} 
\begin{document} 
\maketitle

\begin{abstract}
The broad energy response, low electronic read noise, and good energy resolution have made X-ray Charge-Coupled Devices (CCDs) an obvious choice for developing soft X-ray astronomical instruments over the last half century. They also come in large array formats with small pixel sizes which make them a potential candidate for the next generation astronomical X-ray missions.  
However, the next generation X-ray telescopic experiments propose for significantly larger collecting area compared to the existing observatories in order to explore the low luminosity and high redshift X-ray universe which requires these detectors to have an order of magnitude faster readout. In this context, the Stanford University (SU) in collaboration with the Massachusetts Institute of Technology (MIT) has initiated the development of fast readout electronics for X-ray CCDs. 
At SU, we have designed and developed a fast and low noise readout module with the goal of achieving a readout speed of 5 Mpixel/s. 
We successfully ran a prototype CCD matrix of 512 $\times$ 512 pixels at 4 Mpixels/s.  
In this paper, we describe the details of the readout electronics and report the performance of the detectors at these readout speeds in terms of read noise and energy resolution. In the future, we plan to continue to improve performance of the readout module and eventually converge to a dedicated ASIC based readout system to enable parallel read out of large array multi-node CCD devices. 
\end{abstract}


{\noindent \footnotesize\textbf{*}Tanmoy Chattopadhyay,  \linkable{tanmoyc@stanford.edu} }

\begin{spacing}{1}   

\section{Introduction}
\label{sect:intro} 
X-ray CCDs \cite{Lesser15_ccd,gruner02_ccd} have been the workhorse of soft X-ray instrumentation for more than three decades. The small pixel size, very low electronic read noise, good energy resolution and broad energy response (of up to tens of keV) have led to X-ray CCDs being employed as sensitive soft X-ray spectro-imagers across many applications, including space missions. X-ray astronomy satellites, including ASCA, Chandra, XMM-Newton, Swift, Suzaku, Hitomi, and AstroSat, utilizing CCD based spectro-imagers have been extremely successful. In particular Chandra, with its high angular resolution optics, has been able to utilize fully the small pixel sizes of CCDs to provide ground breaking high resolution imaging of the X-ray cosmos, from black holes and neutrons stars to supernovae and galaxy clusters.  

The next generation of astronomical X-ray missions will aim to expand on the success of Chandra by combining high angular resolution with large collecting area and a wide field of view, to probe deeper into the low luminosity and high redshift X-ray universe. For example, the Lynx Observatory \cite{gaskin15_lynx}, proposed to NASA's Astro2020 Decadal Survey, has $\sim$30 times the collecting area of Chandra and would provide an unprecedented capability to observe, for example, the birth of supermassive black holes and how these systems go on to shape the galaxies around them. Only slightly less ambitious than Lynx, the Advanced X-ray Imaging Satellite (AXIS \cite{mushotzky2019_axis}) is an example probe-class X-ray mission concept, sharing much of the technology and many of the science goals. Astro2020 recommended that NASA proceeds with technology maturation for Lynx and move immediately ahead with ``an X-ray mission designed to complement the European Space Agency (ESA) Athena mission'' as a top priority for a new line of probe class missions.

To exploit fully the science potential provided by their exquisite X-ray optics, missions like AXIS and Lynx will require low noise, small pixel detectors filling their focal planes. However, to overcome source pile-up and saturation effects \cite{lumb00_pileup_xmm}, and to minimize the detrimental impact of the charged particle background on observations of faint, diffuse sources \cite{dan20_bkg}, these detectors must be an order of magnitude faster than existing X-ray CCDs.

The past decade has seen significant progress in the development of fast readout, small pixel size, and low power X-ray detectors (see \cite{falcone18_HXDI} for an overview). Active Pixel Sensors (APS) such as X-ray Hybrid CMOS detectors \cite{bai08,hull17,chattopadhyay18_HCDoverview,hull18_small_pixel}, the depleted field effect transistor (DEPFET) sensors being developed for the ATHENA Wide-Field Imager (WFI, \cite{norbert16_wfi}), and the Silicon-On-Insulator (SOI) technology \cite{SOI18} promise to fulfill some of these requirements. Most of the requirements are also met with the proven technology of X-ray CCDs, with the aforementioned of exception of readout speed. Recently, however, papers \cite{bautz18,bautz19,bautz20} have reported significant progress in the development of low power, high speed X-ray CCDs. In particular, MIT Lincoln laboratory (MIT-LL) is developing CCDs with fast, low-noise on-chip amplifiers 
and gate electrodes formed with a single-level polysilicon process that enable low power operation of the CCDs. Stanford University is collaborating with MIT and MIT-LL to develop fast readout electronics to characterize these CCDs and optimize their noise performance. A detailed overview of the CCDs can be found in \cite{bautz18,bautz19,bautz20}. Here we describe the readout module designed to run the CCDs at high readout speeds, and discuss the spectroscopic and noise performance. 

The structure of this paper is as follows. In Section~\ref{detector}, we provide a brief overview of the MIT-LL digital X-ray CCDs. This is followed by a summary of the characterization test stand and the readout electronics module in Sec. \ref{stand}. We characterized an MIT-LL DCCD, design CCID85B, at readout speeds of 2, 3 and 4 MHz, respectively. The characterization test results are discussed in Sec. \ref{results}. The results are very promising, prompting further optimization of the readout module to characterize the CCDs at an even higher readout speed of 5 MHz. Supported by an ASIC-based, highly parallel signal processing chain, our technology will enable the operation of X-ray CCDs at the high frame rates required by next generation X-ray astronomical missions. 

\section{An Overview of the p-JFET X-ray CCDs}\label{detector}
\begin{figure}
    \centering
    \begin{subfigure}{.36\textwidth}
    \includegraphics[width=\linewidth]{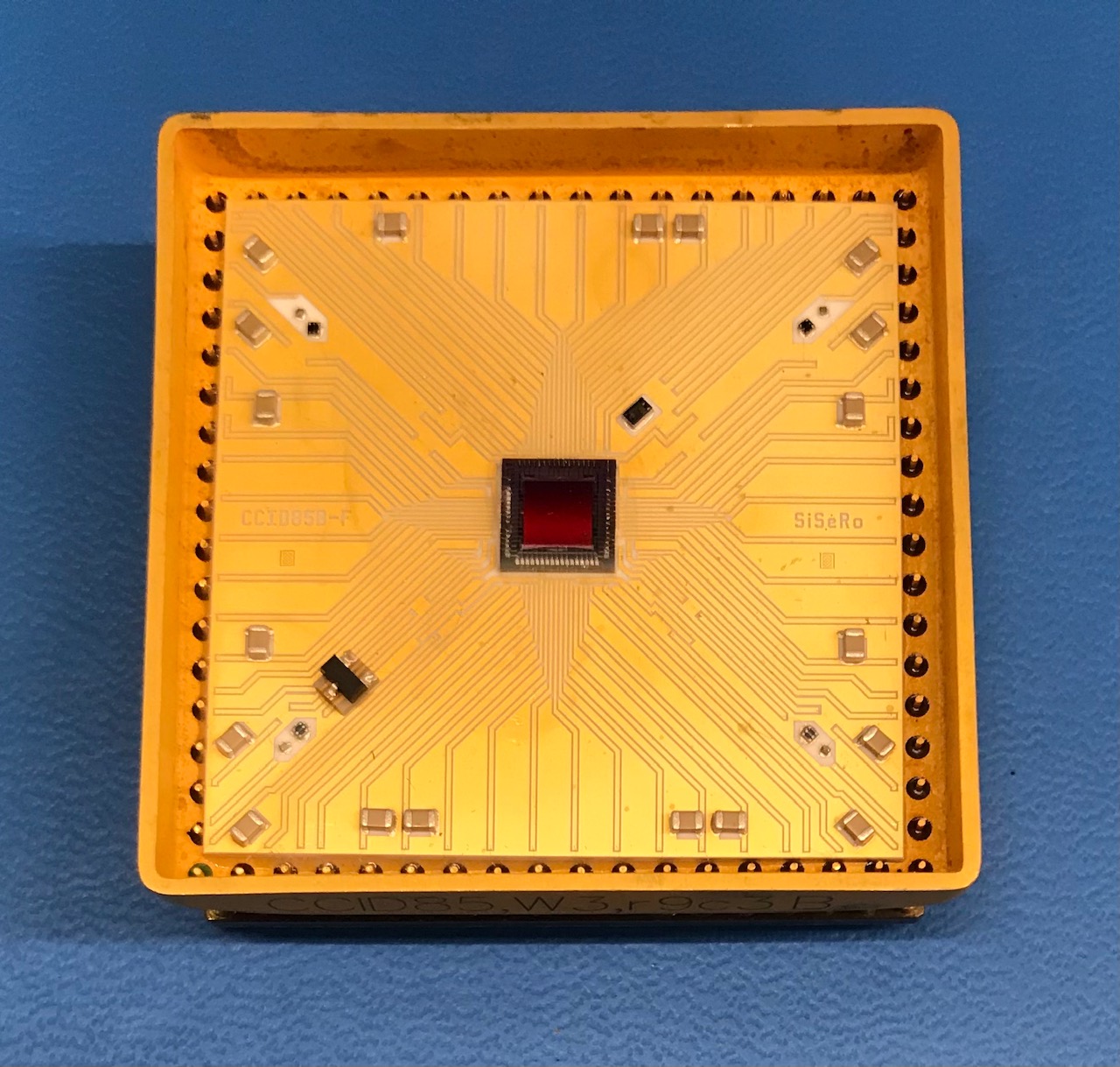}
    \caption{}
    \end{subfigure}
  \begin{subfigure}{.55\textwidth}
     \includegraphics[width=\linewidth]{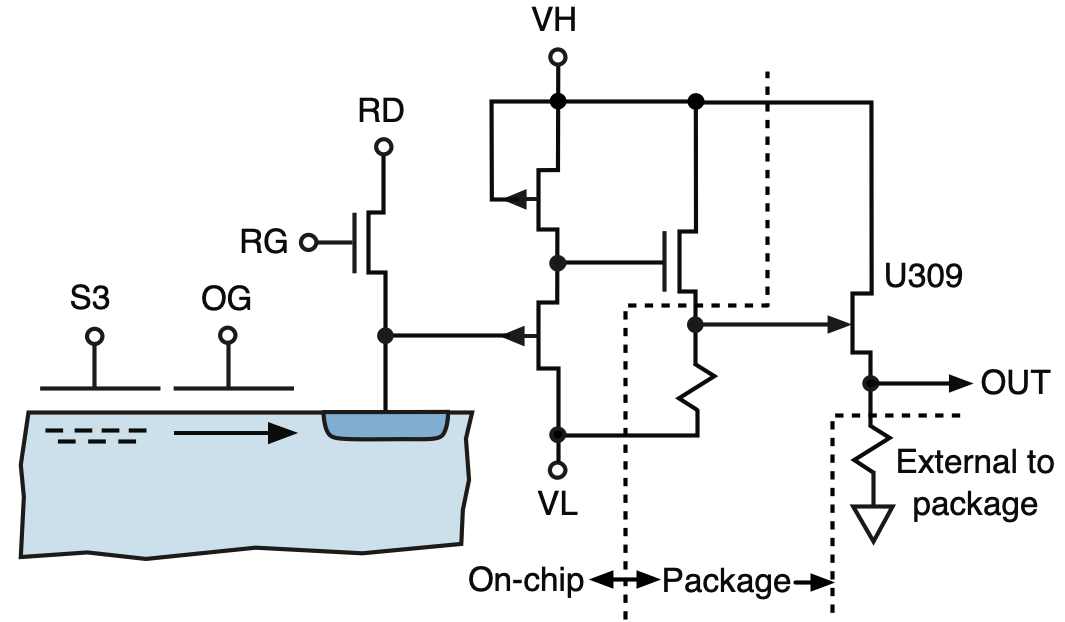}
    \caption{} 
  \end{subfigure}%
    \caption{Prototype CCID85 X-ray CCDs from MIT Lincoln laboratory. (a) One CCID85 sensor in its package that was used for testing. (b) Output stage of the device $-$ a P-JFET floating diffusion amplifier. See text for more details.}
    \label{ccid85}
\end{figure}

The prototype CCID85B CCD detectors used for testing are fabricated in an n-channel, low-voltage, single-poly process and use fast, low-noise on-chip amplifiers. 
Figure \ref{ccid85}a shows a prototype CCID85 device package. The sensitive volume is $\sim$ 4 mm $\times$ 4 mm in size with a pixel array of 512 $\times$ 512, where the pixel side dimension is 8 $\mu$m. 
Conventional CCDs use buried-channel MOSFETs as the sense transistor, however, to meet the frame rate and noise performance requirements of the next generation X-ray astronomy missions, these devices employ a p-JFET source follower as the first stage followed by a buried-channel MOSFET at the second stage to increase the bandwidth of the amplifier. Figure \ref{ccid85}b shows the schematic of the output stage amplifier. There can also be an optional third stage with a U309 n-JFET source follower within the package. However, the device we used for testing was not equipped with a U309 source follower.  
The CCID85 prototype detectors are front-illuminated devices. S$_3$ and OG in the figure stand for the last serial clock at the output stage and the Output Gate respectively. The OG transfers the charge packet from S$_3$ to the floating diffusion gate (FD) where the charge modulates the output of the p-JFET amplifier. The charge packet is eventually drained to the RD, which is biased at high DC potential, using a reset clock (RG). More details on the charge transfer can be found in Sec. \ref{method}.

\section{Characterization Test Stand}\label{stand}
\begin{figure}
    \centering
    \begin{subfigure}{.59\textwidth}
    \includegraphics[width=\linewidth]{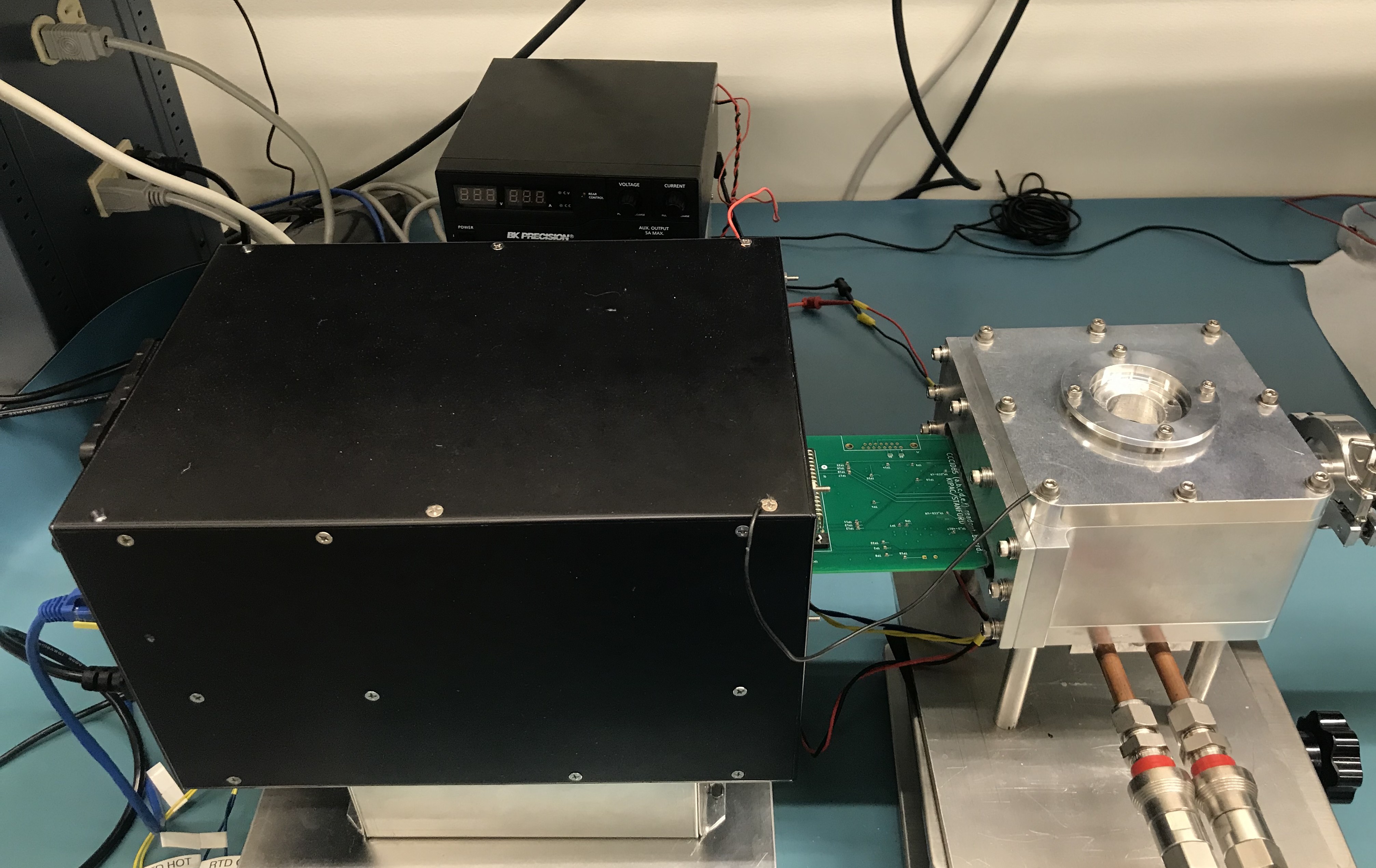}
    \caption{}
    \end{subfigure}
  \begin{subfigure}{.402\textwidth}
     \includegraphics[width=\linewidth]{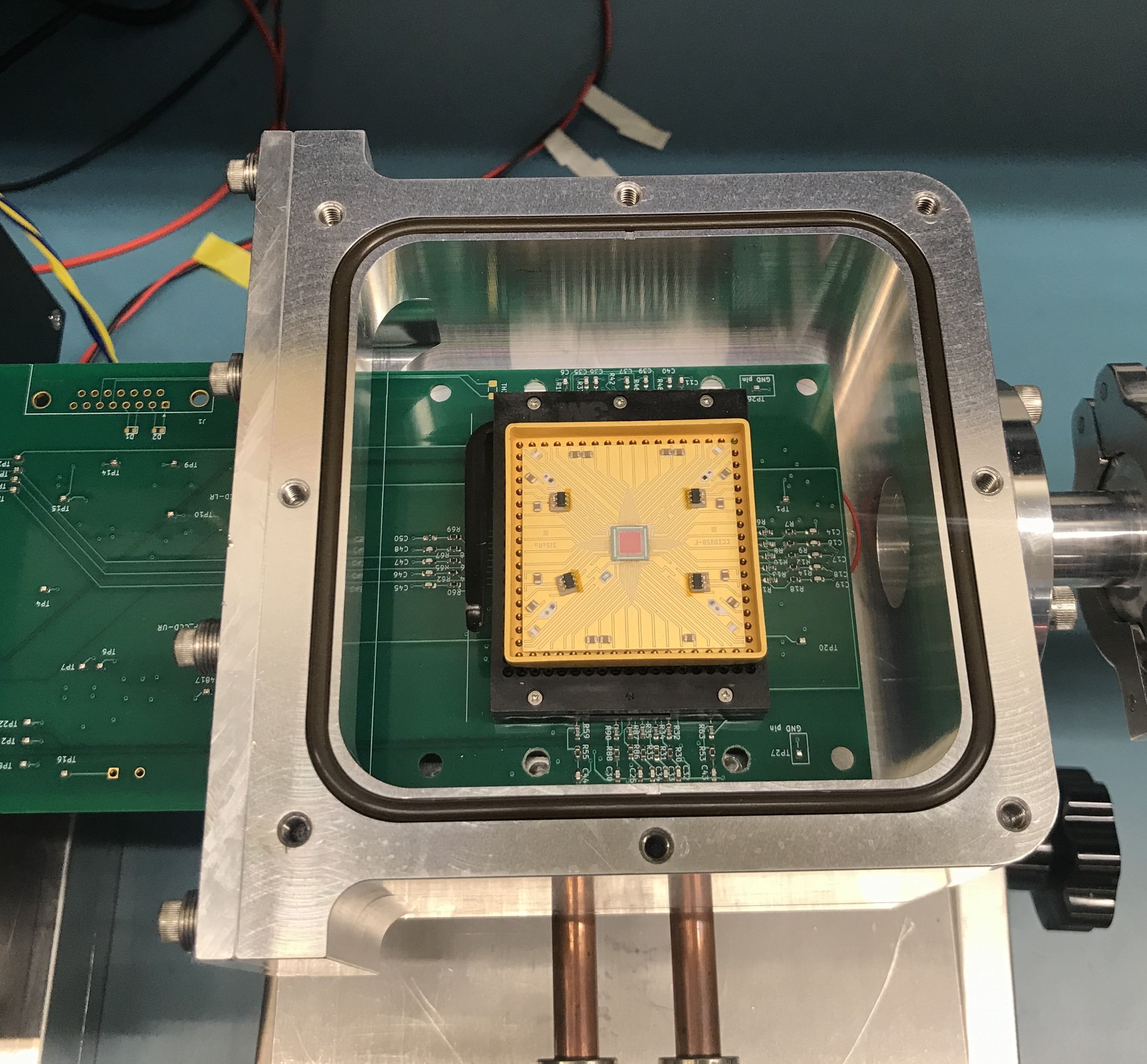}
    \caption{} 
  \end{subfigure}%
    \caption{(a) The characterization test stand with a CCID85 device mounted inside the chamber. (b) Inner view of the chamber and the test device. The detector is clamped against the thermal cold block with a metal plate from the top for cooling (not shown in the figure). For more details of the test set up and its components, see \cite{Chattopadhyay20_spie}.}
    \label{expt_setup}
    \end{figure}

Figure \ref{expt_setup} shows the complete experiment setup. 
The characterization test stand and its different components have already been discussed in details in our previous paper \cite{Chattopadhyay20_spie}. Therefore, here we only give a brief description of the test stand followed by a discussion on the readout module.

\subsection{Detector Housing}
The CCD housing, shown in Fig. \ref{expt_setup}a, is compact with dimensions of 13 cm $\times$ 15 cm $\times$ 6.5 cm. 
One of the side flanges has a narrow slot to epoxy the CCD preamplifier board such that nearly half of the board with a CCD mounting socket is inside the chamber (see Fig. \ref{expt_setup}b). The other half is outside the chamber and connects to an Archon CCD controller\footnote{http://www.sta-inc.net/archon/}, the black box shown in Fig. \ref{expt_setup}a (details of the Archon controller will be discussed below).
An aluminum cold block fits through a square hole on the vacuum side of the board.
For the detector mounting socket, we machine a 361 (19 $\times$ 19) positions PGA ZIF socket, procured from 3M, with a square hole (with finally 68 output pins in total along the boundary of the socket) such that the device package is in thermal contact with the cold block. 
The cold block is epoxied to a 2-stage thermo-electric cooler (TEC) installed on the bottom flange of the chamber to cool  the detectors to low temperatures. We employ a water cooled plate on the other side of the bottom flange to dissipate the heat from the TEC. In addition, a PID loop regulates the current to the TEC and thereby controls the temperature of the CCD. With the current setup, the detectors can be cooled down to -25$^\circ$C with temperature fluctuations constrained to within 0.2$^\circ$C.   
An X-ray entrance window, made of a 500 $\mu$m thick beryllium disc with a diameter of 3.8 cm is installed in the top flange directly above the detector.  
The window transparency is around 95 \% for transmission of 5.9 keV photons, enabling effective use of the Manganese (Mn) k$_{\alpha}$ line (5.9 keV) of a standard Iron-55 ($^{55}$Fe) radioactive source to characterize the detectors. The flange opposite to the readout board is connected to a pump station. 

\subsection{Readout Module}\label{readout}
The readout module comprises a preamplifier board and the Archon CCD controller. The primary function of the board is to amplify the CCD output signal and convert it to a differential signal for the ADCs in the Archon controller.      
To achieve faster readout speeds, a wider bandwidth is required. Concurrently, noise increases with the square root of bandwidth. This, in turn, imposes restrictions on the choice of amplifiers for the readout circuit. 
The CCD on-chip amplifier with a transconductance (g$_m$) of 166 $\mu$S (g$_m$ typically ranges between 166 $\mu$S and 1000 $\mu$S for these devices) results in the CCD output referred (ORN) spectral noise density of around 10 nV/$\sqrt{Hz}$ at room temperature. To keep the overall noise of the system as low as possible, the input referred noise (IRN) of the readout module is desired to have a value lower than or at least comparable to 10 nV/$\sqrt{Hz}$. Therefore, for the readout circuit, we consider amplifiers with low voltage and current noise densities, large bandwidths, and low input capacitances. The low input capacitance forms the RC time constant of the circuit where the C is the total input capacitance of the circuit and R is the reciprocal of the transconductance (g$_m$) of the CCD on-chip amplifier. 

\begin{figure}
    \centering
    \begin{subfigure}{1\textwidth}
    \includegraphics[width=\linewidth]{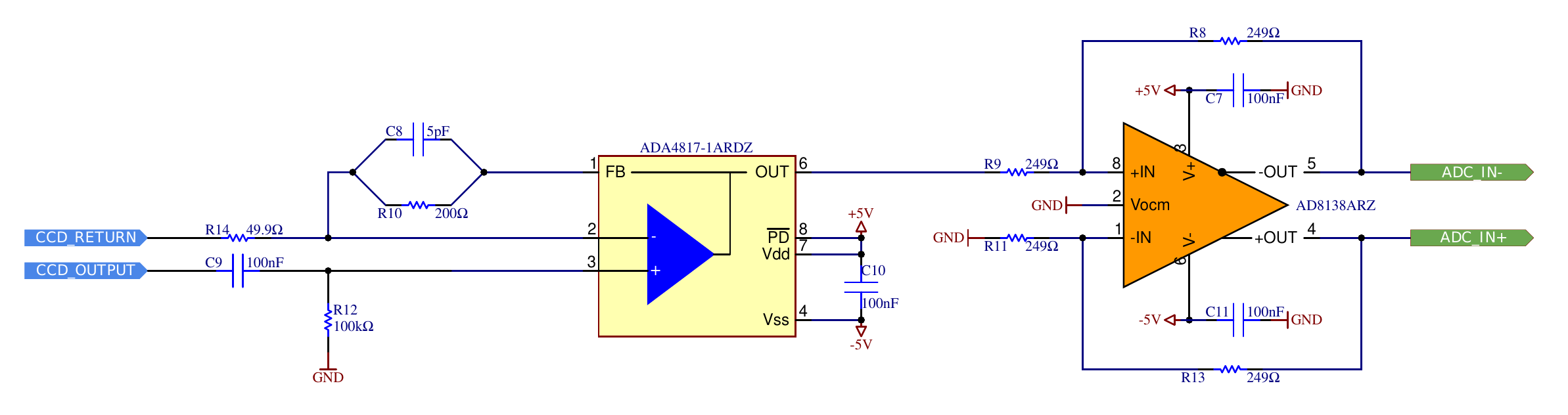}
    \caption{}
    \end{subfigure}
  \begin{subfigure}{.52\textwidth}
     \includegraphics[width=\linewidth]{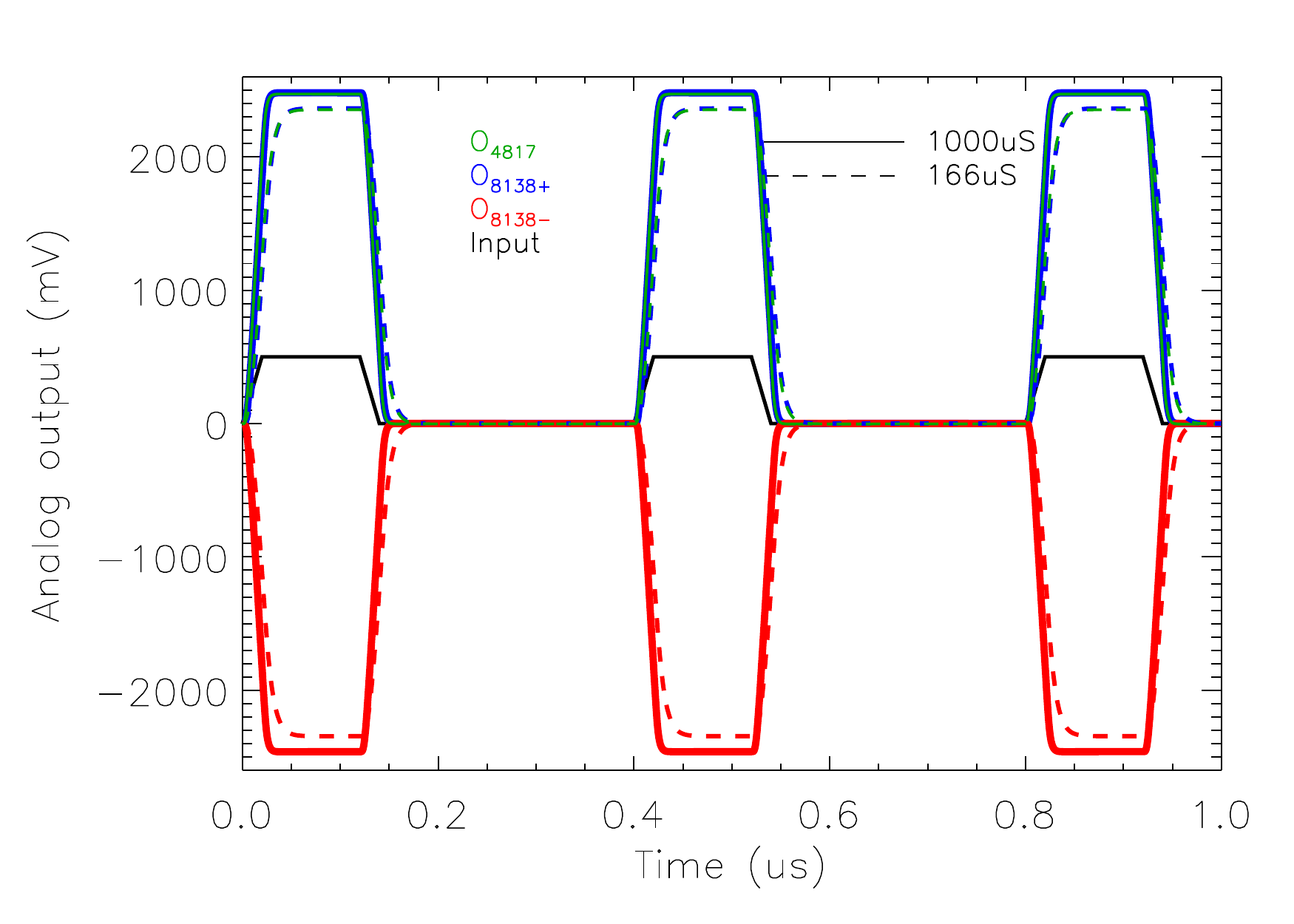}
    \caption{} 
  \end{subfigure}%
  \begin{subfigure}{.52\textwidth}
     \includegraphics[width=\linewidth]{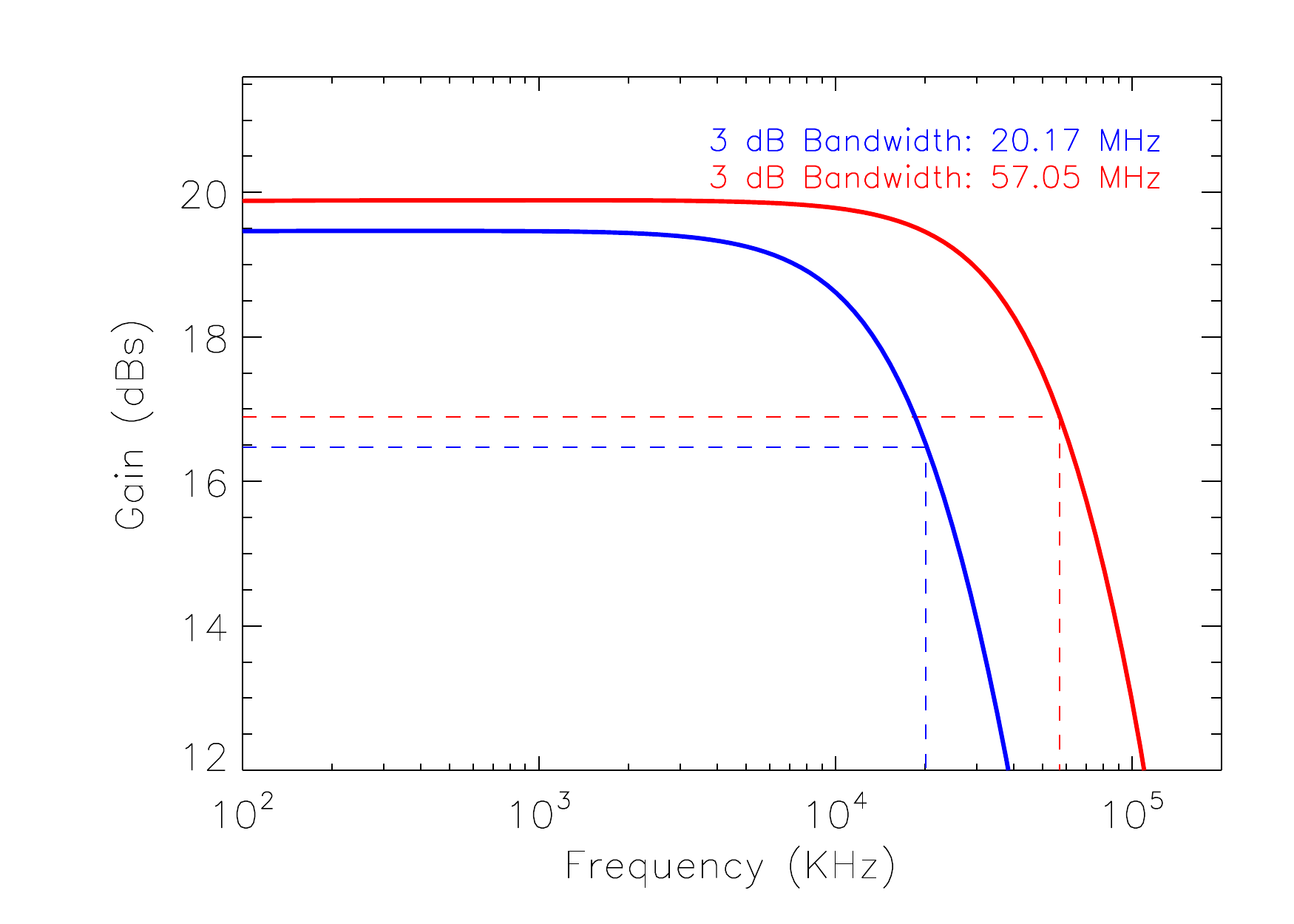}
    \caption{} 
  \end{subfigure}%
    \caption{(a)Schematic of the preamplifier circuit to read out the CCD analog output. It uses a single-ended operational amplifier (ADA4817) in a non-inverting configuration as the first stage, which is followed by a fully differential ADC driver (AD8138). See text for more details. (b) Transient simulation results for the preamplifier circuit in LTspice simulator. Output from different stages of the preamplifier chain is shown in different colors for a pulsed input signal with a repetition rate of 2.5 MHz. (c) Predicted bandwidth for two different CCD on-chip amplifier transconductance values (g$_m$=166 $\mu$S and 1 mS). For higher g$_m$, the board is expected to provide $\sim$57 MHz of bandwidth.}
    \label{preamp_schematic}
\end{figure}

Figure \ref{preamp_schematic}a shows the schematic of the preamplifier circuit. 
The circuit uses a single-ended operational amplifier in a non-inverting configuration  (ADA4817 from Analog Devices\footnote[3]{https://www.analog.com/en/products/ada4817-1.html$\#$product-overview}) as the first stage, followed by a differential ADC driver (AD8138 from Analog Devices,\footnote[4]{https://www.analog.com/en/products/ad8138.html$\#$product-overview}). Such a configuration produces a fully differential signal that can be directly interfaced to the ADCs. The ADA4817 provides extremely low input capacitance of $\sim$1.3 pF and therefore is best suited for this type of application.
The overall bandwidth of the circuit is nearly independent of the second stage. The ADA4817 and the AD8138 both have low voltage noise densities ($<$5 nV/$\sqrt{Hz}$) which ensures a low ORN from the setup. 

We simulated the preamplifier circuit in the LTspice simulator\footnote[5]{https://www.analog.com/en/design-center/design-tools-and-calculators/ltspice-simulator.html}, results of which are shown in Fig. \ref{preamp_schematic}b and c. 
The simulations were done for two different resistor values at the input of the first stage $-$ 6 k$\Omega$ and 1 k$\Omega$ representing transconductances of the CCD on-chip amplifier of 166 $\mu$S and 1 mS, respectively. 
Figure \ref{preamp_schematic}b shows the signal output from the first stage, ADA4817 (shown in green), inverting and non-inverting output of the ADC driver, AD8138 (shown in red and blue respectively) for a pulsed input signal of 500 mV amplitude and 2.5 MHz repetition rate (shown in black).  
The overall bandwidth is shown in Fig. \ref{preamp_schematic}(c). The upper cutoff is around 21 MHz for the 6 k$\Omega$ input resistor whereas for the 1 k$\Omega$ resistance, it is around 57 MHz due to the lower RC time constant of the circuit. 
The effect of the RC time constant can also be seen in (b), where the outputs of the first stage, ADA4817 and the differential driver, are clearly seen to be slowed down for the higher resistance value (dashed lines). The noise simulations predict around 15 nV/$\sqrt{Hz}$ noise density at the input of the readout module including the noise of the CCD on-chip amplifier.

A 50-pin connector, on the other side of the board, connects the preamplifier board to the Archon controller. An interface board routes the clock and biases from the internal Archon modules to the CCD through the preamplifier board and routes the analog signals from the preamplifier board to the differential ADCs of the Archon. We implement filtering circuitry for the clock signals and bias voltages in the preamplifier board.

The Archon controller \cite{archon14}, procured from Semiconductor Technology Associates, Inc (STA), is an FPGA-based modular high performance CCD controller, which provides the necessary bias and clock signals to the CCD, digitizes the CCD output, and extracts the signal charge information from the digitized CCD waveform.
The Archon provides a total of 12 slots for the ADC, the clock driver, the bias, the heater, or other custom modules, with up to 4 ADC modules for a total of 16 CCD outputs. 
CCD outputs are digitized by 16-bit 100 MHz ADCs. CCD clocks are generated by 14-bit 100 MHz DACs. The LV bias module provides a total of 30 biases with voltages ranging from -14 V to 14 V. The HV bias module also provides 30 total biases in a 0$-$31 V range. The Archon receives configuration information about the CCD, e.g. bias signals, clock signals, clocking sequence, sampling of digitized waveform to generate an image, from a host PC and returns the status and image data to the host PC via a gigabit Ethernet connection. 


\section{Experiment Methodology and Data Analysis}\label{method}
The primary goals of the tests are: running the detectors at a high readout speed and investigate the limiting factors in the speed performance of the detectors and the readout module, and evaluation of the noise and spectral performance of the detectors at those readout speeds.  

\begin{figure}
    \centering
    \includegraphics[width=0.7\linewidth]{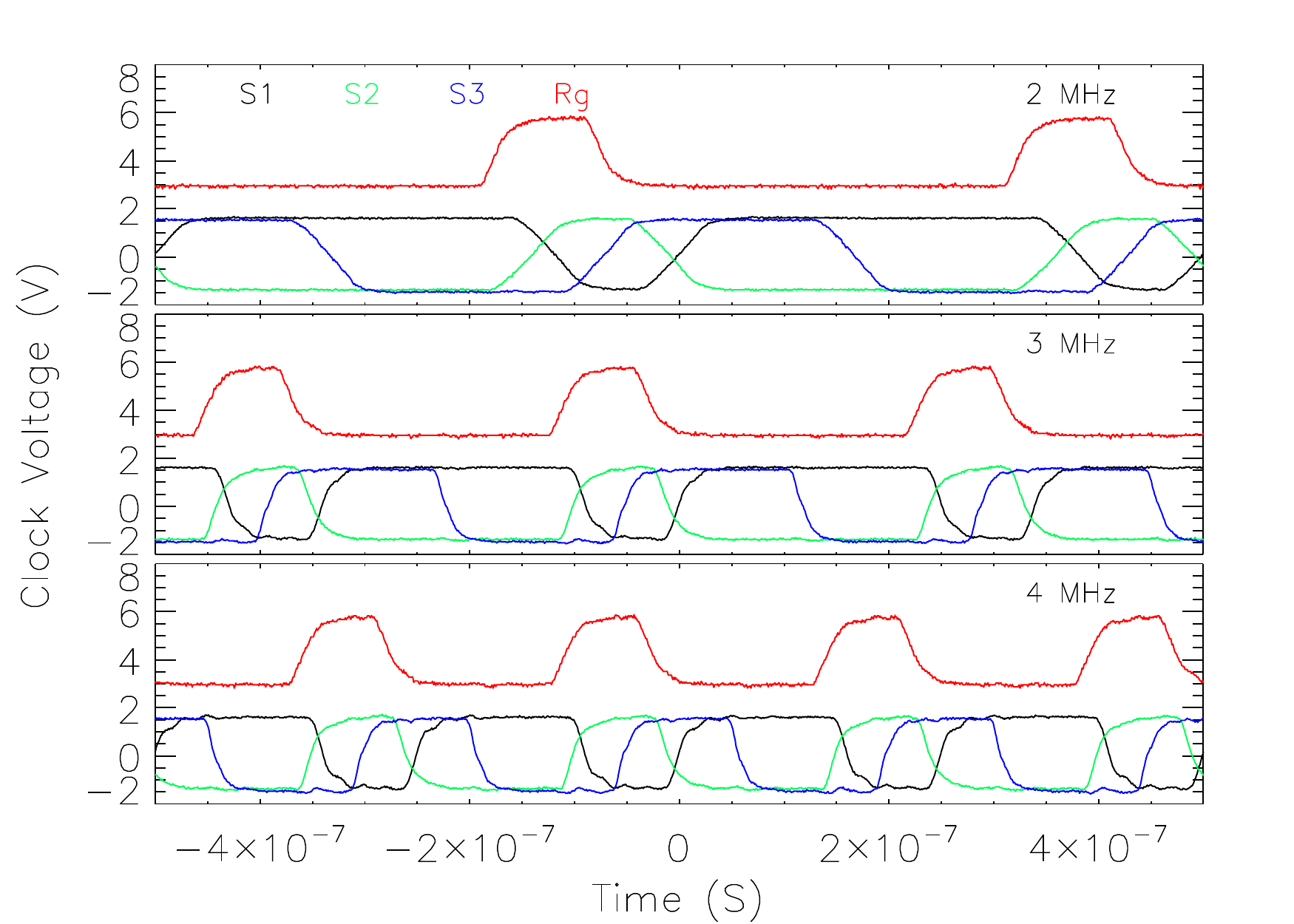}
    \caption{Serial clocks (S$_1$, S$_2$, S$_3$ shown in black, green and blue respectively) and the reset clock (RG show in red) from oscilloscope at different readout speeds -- 2 MHz (top), 3 MHz (middle) and 4 MHz (bottom) respectively. The charge is transferred to the output gate when S$_3$ is low and then read out by the on-chip amplifier.}
    \label{expt_clocks}
\end{figure}

The device was cooled down to 250 K (-23$^\circ$C). Since the test detectors are front-illuminated devices, the dark current was negligible at 250 K. We characterized the detector at readout speeds of 2, 3 and 4 Mpixel/s, respectively. The readout speed is controlled by changing the time length of the serial clocks (S$_1$, S$_2$, S$_3$) and the reset clock (RG) in the timing script of the Archon Configuration File (ACF). Figure \ref{expt_clocks} shows the serial and reset clocks of 2, 3 and 4 MHz in the top, middle, and bottom panels, respectively.
The charge is first transferred from the imaging region to the serial register using a three electrode imaging clock sequence running at 400 KHz. The charge is then moved from S$_1$ to S$_2$ and S$_2$ to S$_3$ and finally to the floating diffusion (FD) gate (see Fig. \ref{detector}) using the clock sequence shown in Fig. \ref{expt_clocks}. After the charge is read out by the on-chip amplifiers, RG drains the charge to the high voltage reset drain (RD).  

\begin{figure}
    \centering
    \includegraphics[width=0.7\linewidth]{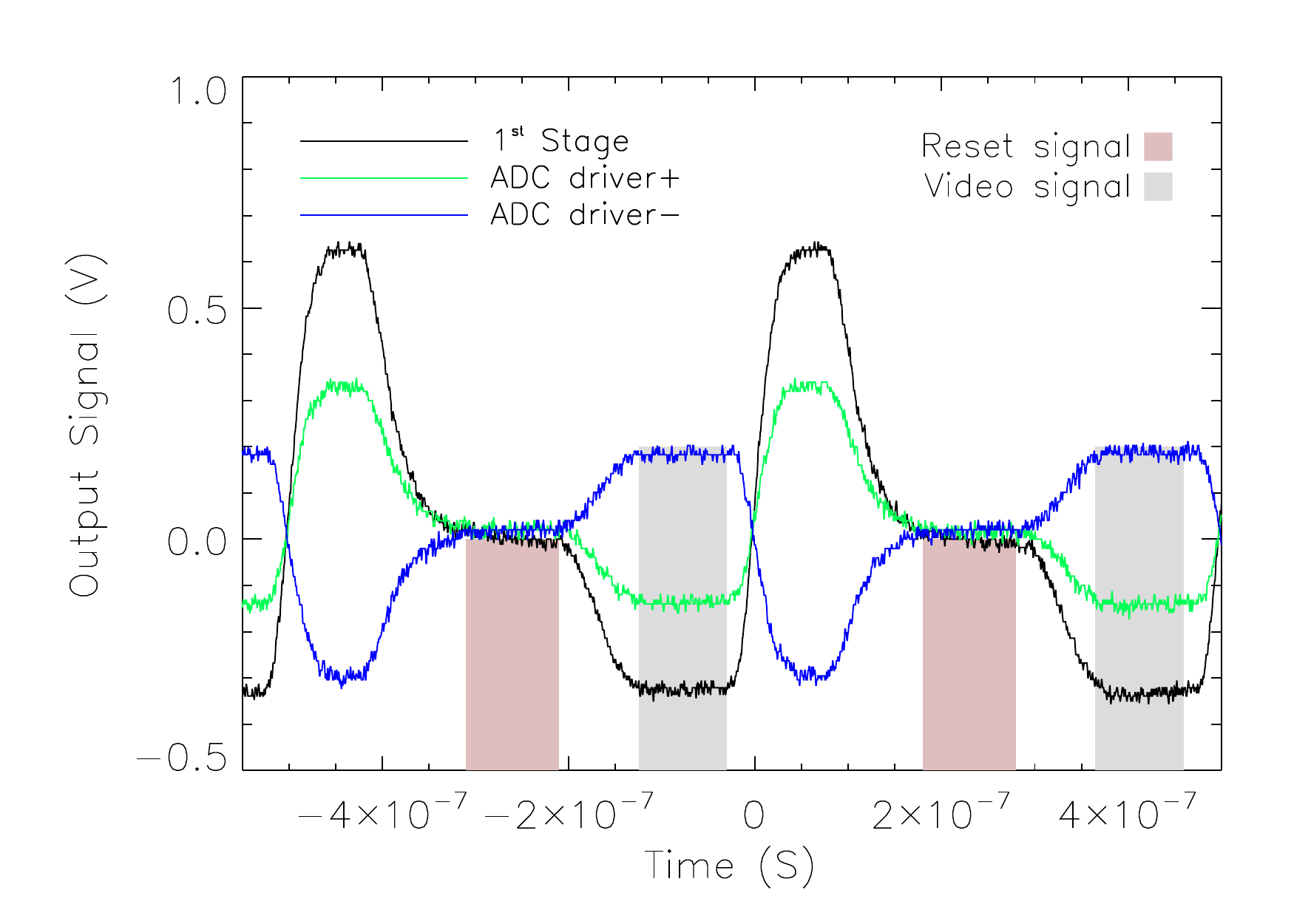}
    \caption{Oscilloscope trace showing the output signal of the 1$^{st}$ stage amplifier (black) and the ADC driver (green and blue) at a readout speed of 2 MHz. The shaded regions represent the reset and video signals which are used for CDS filtering. The difference of the two levels is proportional to the source signal.}
    \label{expt_amplifier}
\end{figure}

Figure \ref{expt_amplifier} shows the CCD output from an oscilloscope after the amplification stage. 
The black line shows the output of the first stage of the pre-amplifier (ADA4817) whereas the blue and the green waveforms are taken at the inverting and non-inverting outputs of the differential ADC driver respectively for a readout speed of 2 MHz.  
Every pixel of the CCD generates such a signal at the output. The initial part of the waveform is formed by the reset pre-charging the floating diffusion and the gate of the p-JFET transistor to a predefined potential. Following the reset, the output settles to a baseline level (see the first shaded region in Fig. \ref{expt_amplifier}). After the charge packet is transferred from the serial register to the floating diffusion (when S$_3$ is low), the p-JFET output will settle to a new voltage level (see the second shaded region in Fig. \ref{expt_amplifier}). 
The signal amplitude is obtained by taking the difference between the two levels (Correlated Double Sampling process or CDS) for each pixel to generate 2D image frames.  

We have developed an IDL (Interactive Data Language) based event processing pipeline for image cleaning, event selection, generation of spectrum, and estimation of noise and spectral resolution. To estimate the read noise of the system, we overclock an extra 50 columns such that for each frame we generate a 562 $\times$ 512 array image. Read noise is estimated from the overclocked region by calculating the standard deviation in the charge distribution of the 50 $\times$ 512 array of pixels.  
To test the overall noise performance of the system and estimate the contribution of the leakage current to the total noise, we collect dark frames at multiple frame integration times (5 ms to 3 seconds) for 2 MHz and 4 MHz readout speeds, respectively.     
An $^{55}\mathrm{Fe}$ radioisotope was used to evaluate the spectral performance of the device for 5.9 keV (Mn K$_\alpha$) and 6.4 keV (Mn K$_\beta$) X-ray photons. We apply bias (from the overclocked regions at 0 ms integration) and dark frame correction to the raw images to generate clean X-ray images. An event selection logic based on a user-defined primary and secondary pixel charge thresholds is applied to the cleaned X-ray images to generate X-ray spectra of different grades. Here we used a primary threshold of 7 times the read noise and a secondary threshold of 2.6 times the read noise to generate the spectra. 

\section{Results}\label{results}
\subsection{Speed Performance}\label{speed}
\begin{figure}
    \centering
    \includegraphics[width=0.7\linewidth]{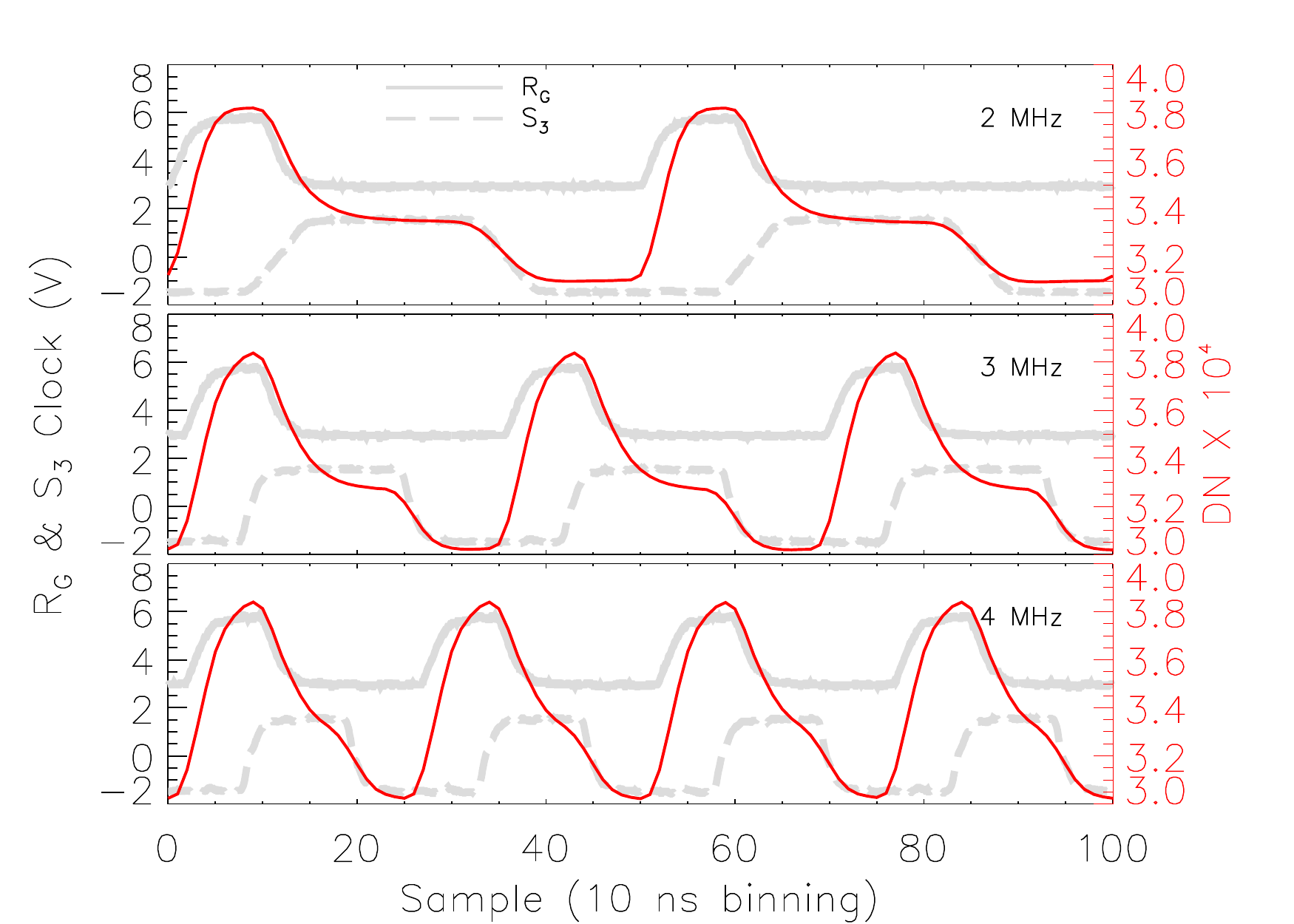}
    \caption{CCD output waveform (red solid line plotted against the right axis) obtained from the Archon controller at multiple readout speeds -- 2 MHz (top), 3 MHz (middle), and 4 MHz (bottom), respectively, along with the reset, RG (solid gray line) and S$_3$ (dashed gray line) clocks. The video signal (see Fig. \ref{expt_amplifier}) starts with charge transfer from S$_3$ to the output gate when S$_3$ is low. The reset flat-top region starts when RG sets to baseline. It is evident that the readout speed of the device is limited by the speed of RG.}
    \label{waveform}
\end{figure}

We characterized the detectors at readout speeds of 2, 3, and 4 Mpixel/s respectively. The readout speed is limited when there are not enough samples either in the baseline or in the signal region or both to perform the CDS filtering. To understand the speed performance and its limitation, we plot the the digitized waveforms obtained from the Archon controller in red solid lines for 2, 3 and 4 MHz in the top, middle and bottom panels of Fig. \ref{waveform}, respectively.   
At 2 MHz, there is a sufficient number of sample points both in the baseline and in the signal region to have an accurate estimate of the charge signals. However, at higher readout speeds, the number of sample points in those regions is too low to perform CDS. Performing CDS is impossible beyond 4 MHz which sets the limiting readout speed for the current setup.  
To understand the limiting factors in the readout speed, we plot the S$_3$, reset clock, RG in Fig. \ref{waveform} (plotted against the left axis in gray lines) along with the corresponding analog waveforms. We see that the extent of the baseline region in the waveform critically depends on the width of the RG clock. Although we program a narrow pulse for the RG, the effective width is considerably longer due to bandwidth limitations of the reset pulse generator. 
Therefore, while the output stage transconductance and loading is already a limiting factor, a faster reset would still result in more time allocation for the signal such that the pixel rate could be increased.
The used RG clocks are generated by 100 MHz DAC in the Archon (10 ns time resolution). In the future, we plan to use an  LVDS clock signal with 1 ns time resolution which are available in some of the Archon models. It might also be possible to use a fast MOSFET clock driver for RG. In the next versions of the readout module, we plan to utilize these ideas to run the detectors at faster readout speeds exceeding 4 MHz. See also the paper \cite{smith18_spie} where some of these ideas in optimization of speed and noise of CCDs have been discussed in detail.         

\begin{figure}
    \centering
    \begin{subfigure}{.49\textwidth}
      \includegraphics[width=1\linewidth]{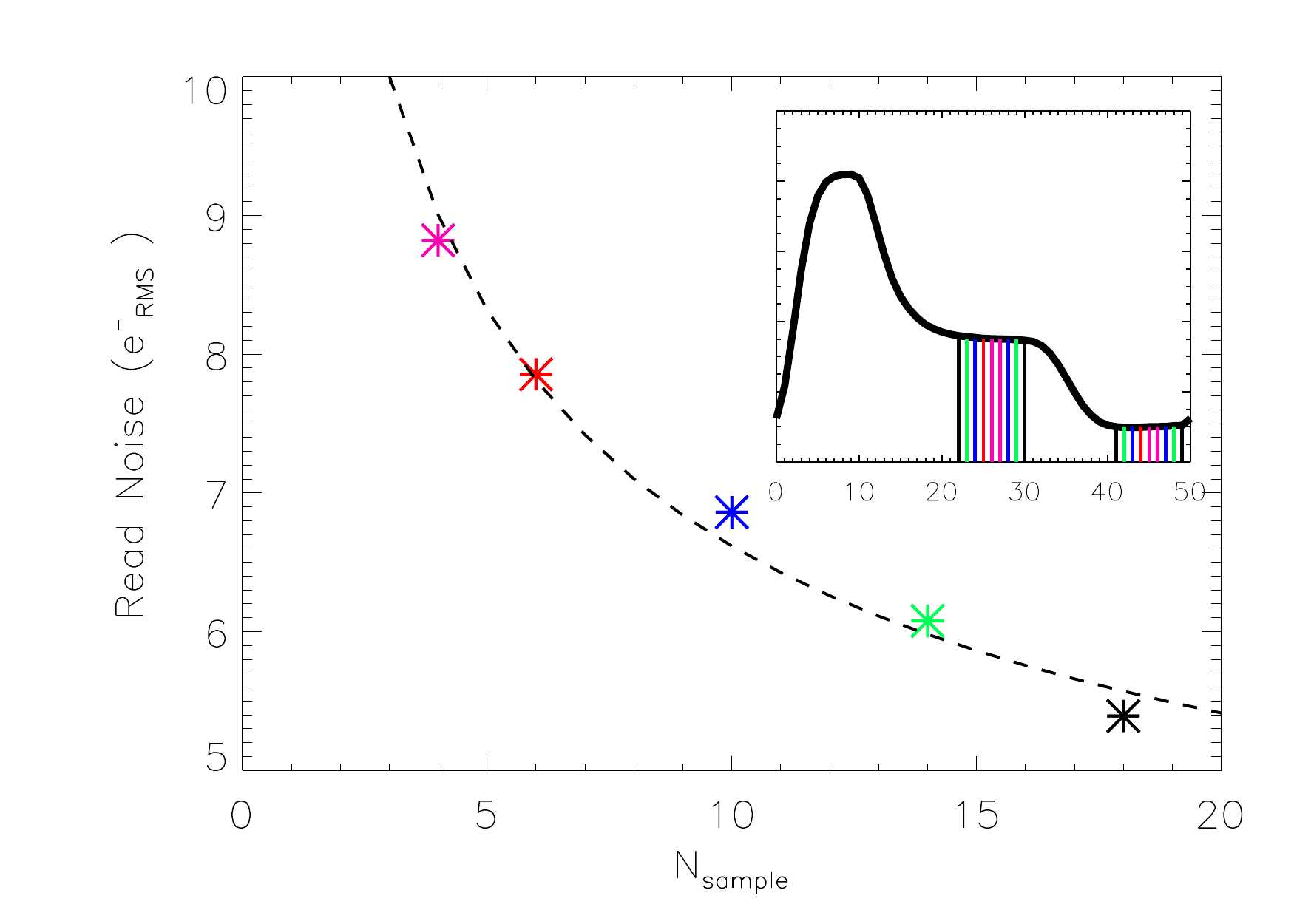}
      \caption{}
    \end{subfigure}
    \begin{subfigure}{.49\textwidth}
      \includegraphics[width=1\linewidth]{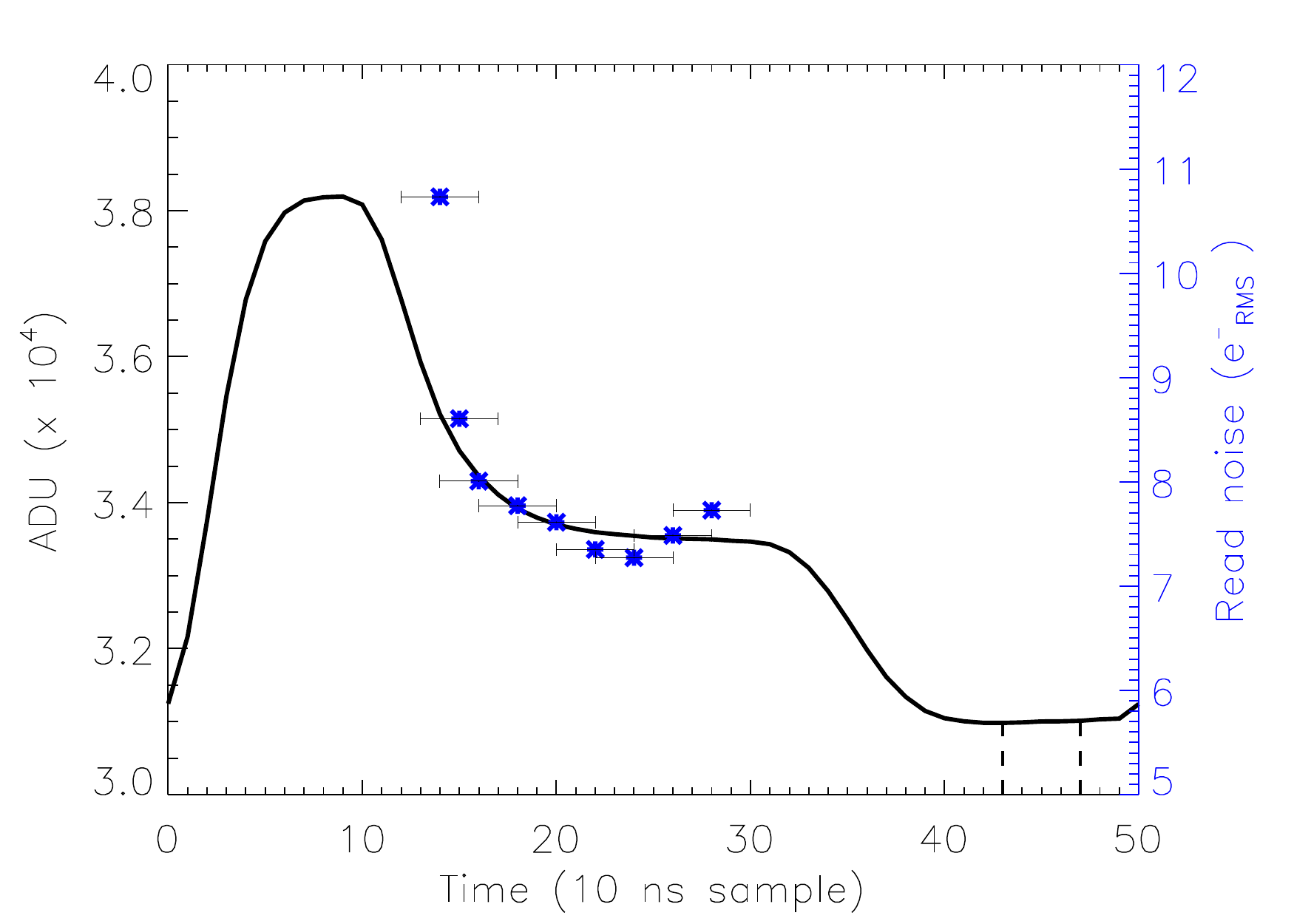}
      \caption{}
    \end{subfigure}
    \caption{(a) Read noise at 2 MHz readout speed as a function of sample size (reset+signal), N$_{sample}$ used in CDS filtering. Different colors stand for different filtering regions shown in the waveform. The dashed black line is a fit to the data points using a function of the form: $A/\sqrt{N_{sample}}+B$. (b) Read noise at a 2 MHz readout speed with different reset or baseline locations for the CDS filtering. The signal region is kept fixed.}
    \label{noise_sample}
\end{figure}

With a faster readout, the system noise is expected to worsen due to two underlying mechanisms that have been identified. First, it is evident from Fig. \ref{waveform} that at higher speeds, the number of sample points to estimate the signal and the baseline is low. Second, the baseline is not entirely settled as it is seen in the third panel of Fig. \ref{waveform}. The second mechanism is in general a form of phase to amplitude modulation where the uncertainty or jitter of the sampling moment causes the voltage signal to be sampled sporadically in the region where the signal or baseline are not settled yet. We quantified the first effect by systematically decreasing the number of samples in the CDS filtering function for both the baseline and the signal region at 2 MHz. Results are shown in Fig. \ref{noise_sample}a where the read noise is seen to increase with a lower number of samples in CDS filtering. Further details of the read noise estimation are discussed in the next section. 
Different colors in Fig. \ref{noise_sample}a represent the width of the filtering regions expressed in number of samples. The trend in noise can be fitted with a function that is proportional to the $1/\sqrt{N_{sample}}$ shown by a dashed black line in Fig. \ref{noise_sample}a. 
In order to quantify the second effect, we keep the the number of samples unaltered while the sampling region is moving along the slope of the reset clock. The effect is shown in Fig. \ref{noise_sample}b where the increase in noise is correlated with the increase in the slope of the waveform. These two mechanism dominate the overall noise of the system at higher readout speeds.

\subsection{Noise Performance}\label{noise}
Read noise is estimated from the distribution of charge in the overclocked region.
The distribution is fitted with a Gaussian to quantify the root mean square (RMS) of the distribution. From the system gain of $\sim$1.2 electrons/ADU (Analog Digital Unit), the read noise is estimated to be around 5.3 $e^{-}_{RMS}$ at 2 MHz, 6.1 $e^{-}_{RMS}$ at 3 MHz, and 7.6 $e^{-}_{RMS}$ at a 4 MHz readout speed. 

\begin{figure}
    \centering
    \includegraphics[width=.9\linewidth]{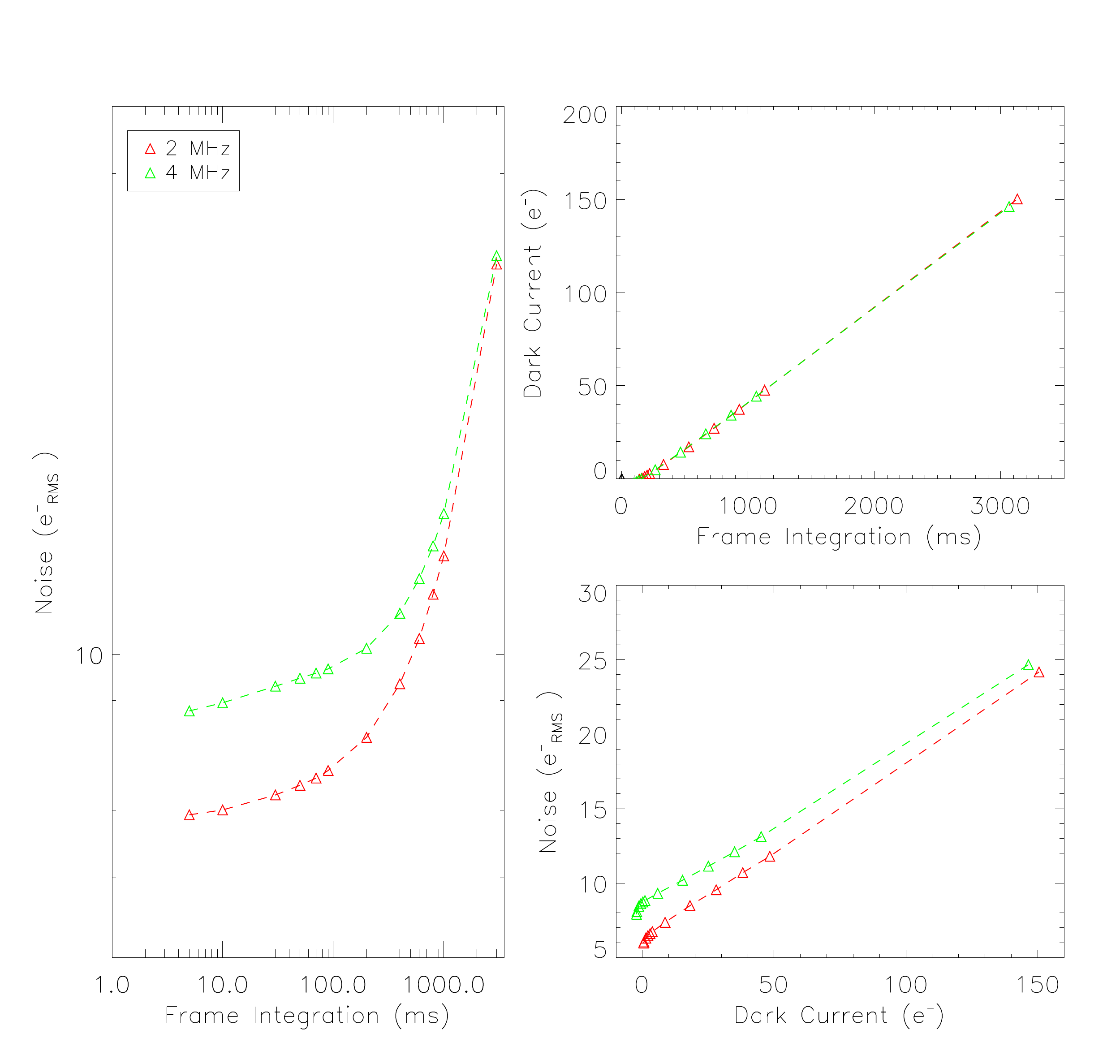}
    \caption{Left: Total system noise as a function of integration time for different readout speeds (red -- 2 MHz and green -- 4 MHz). Right-top: dark current as a function of frame integration time. Right-bottom: system noise as a function of dark current. The total noise can be modelled as the quadrature sum of the read noise and the dark current noise. See text for more details.}
    \label{noise_char}
\end{figure}

The total noise of the system is calculated in the same way by estimating the RMS of charge distribution from the 512 $\times$ 512 pixel imaging area.
Figure \ref{noise_char} (left) shows the variation of the total noise in electrons as a function of frame integration time for 2 MHz (red line) and 4 MHz (green line) readout speeds. 
For shorter integration times, the total noise is dominated by the read noise of the system while for longer integration times the noise is primarily sourced by the leakage or dark current. 
Variation of the mean of dark current (${I_{dark}}$) as a function of integration time along with the variation of the noise with dark current is shown in the right top and bottom panels of Fig. \ref{noise_char} respectively. The mean of the dark current in electrons per millisecond ($e^{-}/ms$) is estimated by fitting a straight line over the dark current plot. It is interesting to note that the noise does not follow a $\sqrt{I_{dark}}$ trend for longer integration times (right bottom panel). The excess noise is possibly due to the hot pixels which are expected to have larger widths in the distribution of charge. The same effect is also seen in the left panel of the figure where noise for an integration time of 3 seconds is larger than expected compared to the dark current noise contribution. The total noise for an integration time of $\leq$1 second can, however, be modeled as a quadrature sum of the electronic read noise estimated from the overclocked region and the $\sqrt{I_{dark}}$.   

\subsection{Spectral Performance}\label{spectral}
\begin{figure}
    \centering
  \begin{subfigure}{.49\textwidth}
    \includegraphics[width=1\linewidth]{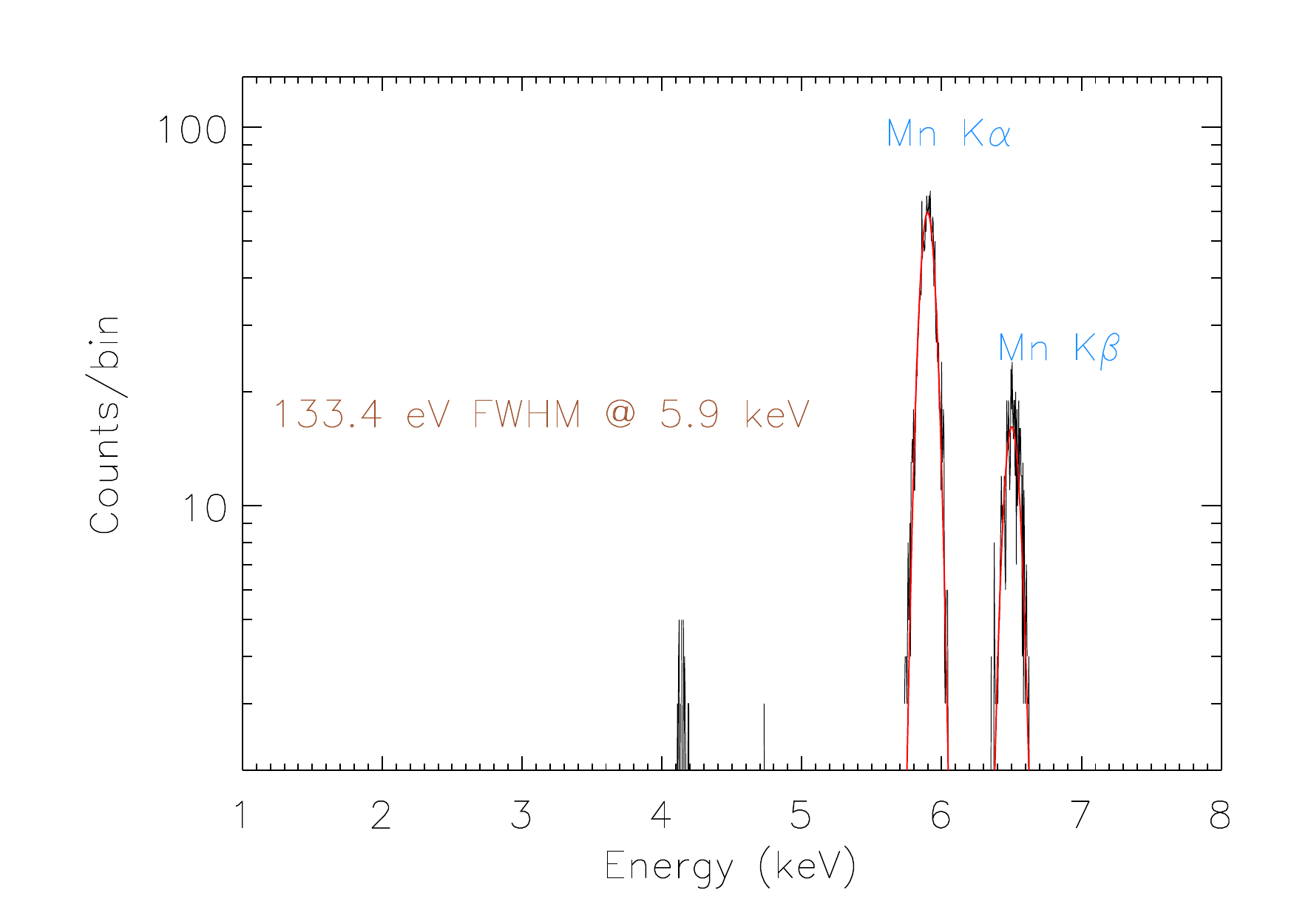}
    \caption{}
    \end{subfigure}
  \begin{subfigure}{.49\textwidth}
    \includegraphics[width=1\linewidth]{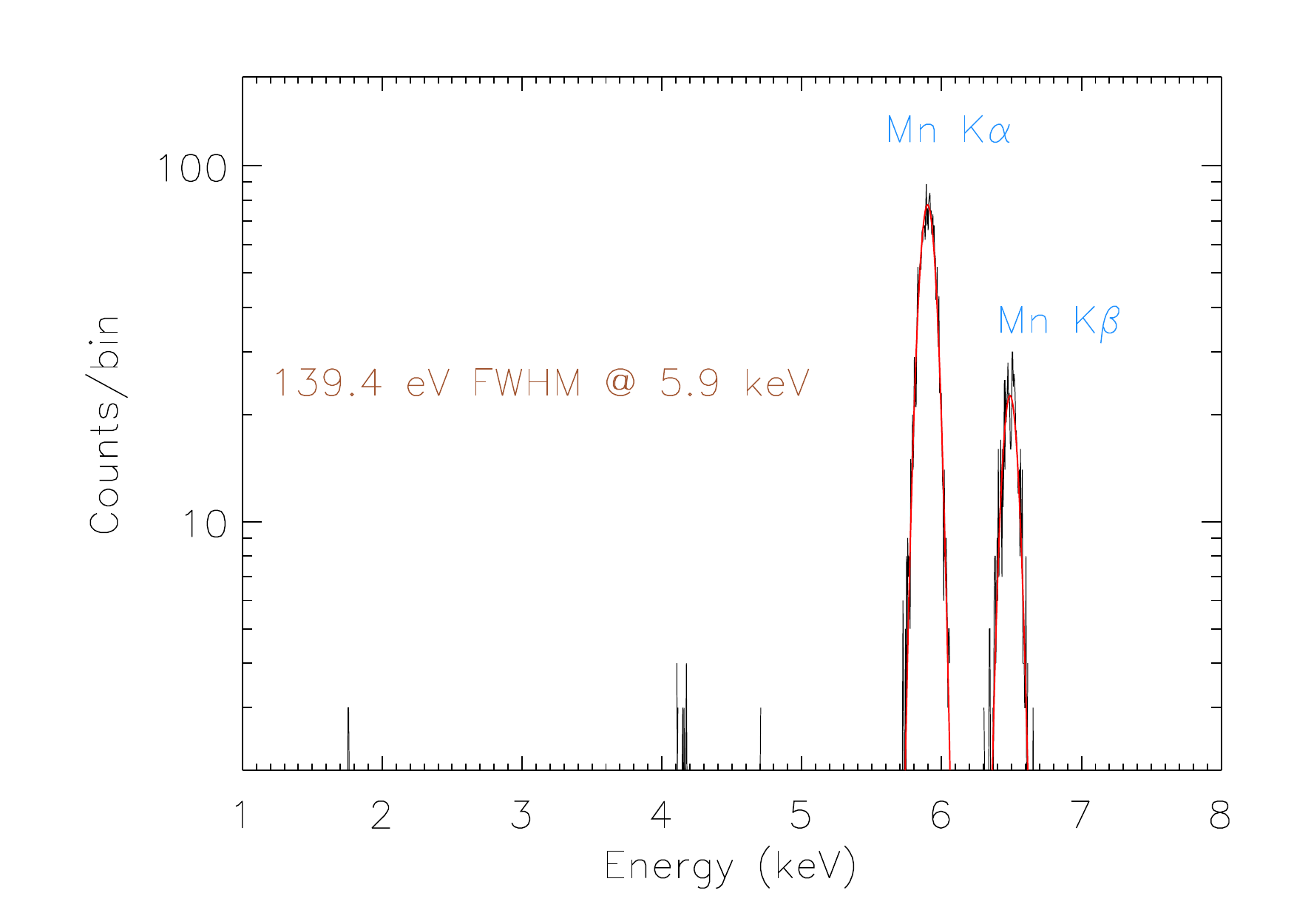}
    \caption{} 
  \end{subfigure} 
  \begin{subfigure}{.49\textwidth}
    \includegraphics[width=1\linewidth]{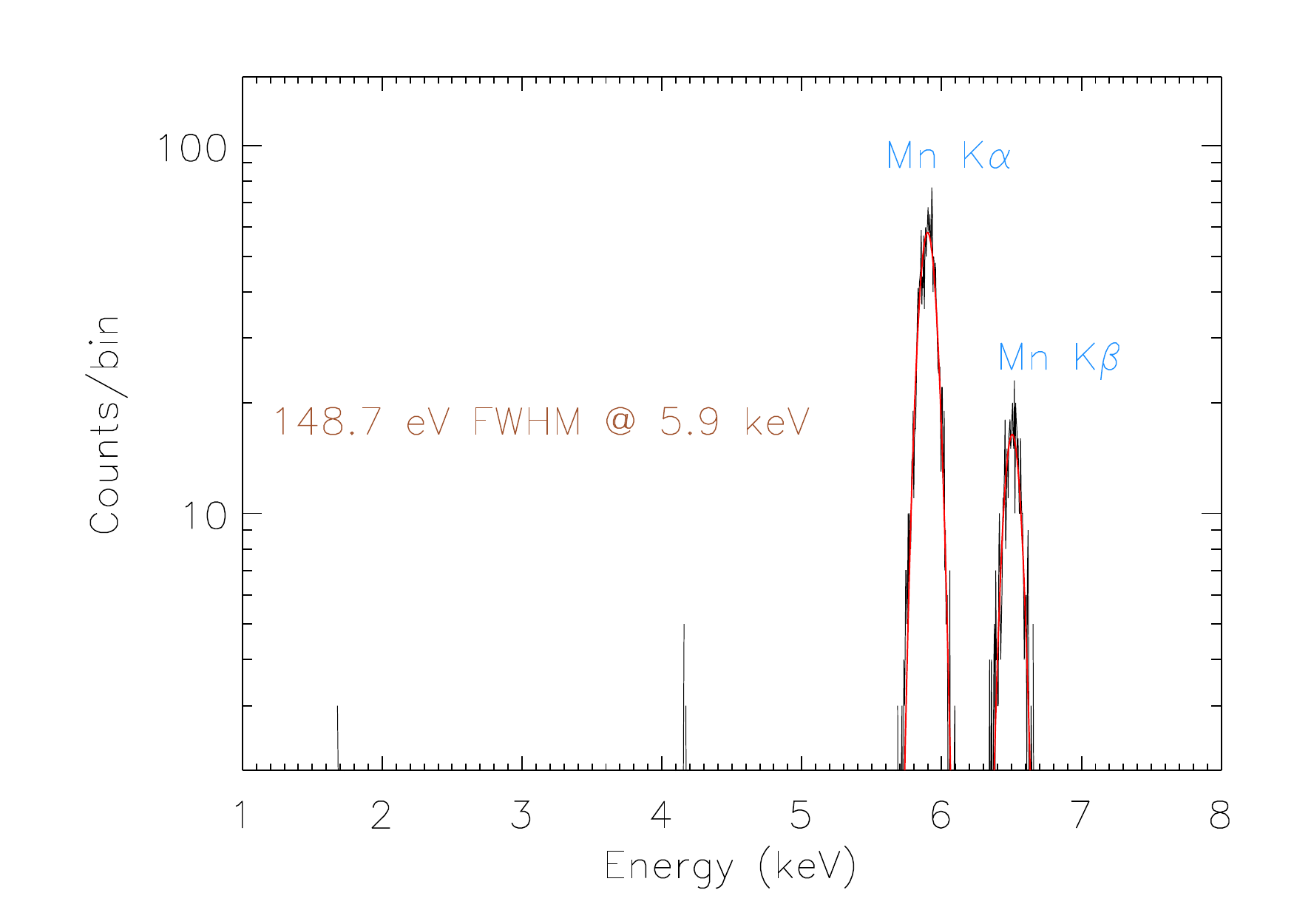}
    \caption{} 
  \end{subfigure}
    \caption{Spectra showing the Mn K$_{\alpha}$ (5.9 keV) and K$_{\beta}$ (6.4 keV) lines from a $^{55}$Fe radioactive source for single-pixel (grade 0) events at (a) 2 MHz, (b) 3 MHz and (c) 4 MHz readout speed. The 5.9 keV and 6.4 keV lines are fitted with two Gaussian functions to estimate the gain and Full Width at Half Maximums (FWHM). The estimated FWHM values at 5.9 keV are 133 eV, 139 eV and 148 eV respectively.}
    \label{spectrum}
\end{figure}

An $^{55}\mathrm{Fe}$ radioisotope was used to generate spectra for the Mn K$_\alpha$ (5.9 keV) and Mn K$_\beta$ (6.4 keV) lines. An event list of 9-pixel islands is generated around each X-ray event. We apply a primary threshold of 7$\times$read noise and a secondary threshold of $\sim$2.6$\times$read noise in ADU to determine the 1-pixel, 2-pixel and multi-pixel events to generate spectra.
Figure \ref{spectrum} shows single pixel or grade 0 X-ray spectrum showing 5.9 keV and 6.4 keV lines at 2 (top panel), 3 (middle panel) and 4 MHz (bottom panel) readout speeds. 
We fit the 5.9 keV and 6.4 keV lines using two Gaussians (shown by red lines) simultaneously to estimate their full width at half maximum (FWHM). We estimate an FWHM of 133 eV for the 5.9 keV line from the single pixel events at 2 MHz readout speed. Because of the small pixel sizes of the device, only a small fraction of the X-ray events is expected to contribute to the single pixel spectrum. We also see a hint of escape peak for the 5.9 keV photons in the spectra. The FWHMs at 3 and 4 MHz readout speeds are estimated to be around 139 eV and 148 eV respectively. From the fitted centroids of the lines and the gain of the readout module, we estimated the conversion gain of the device to be around 28 $\mu$V/electron at 2 MHz.   

\section{Summary and Future Plans}\label{summary}
Stanford University in collaboration with MIT and MIT Lincoln Laboratory is developing low noise readout electronics to characterize the next generation X-ray CCDs which have the potential to fulfill the requirements of low noise and fast readout speed of the future large telescopic astronomical missions.
These test devices, manufactured by MIT Lincoln laboratory, use fast and low noise amplifiers for the output stage.
In this paper, we described the readout electronics module and demonstrated the spectral and noise performances of a prototype X-ray CCD operating at high readout speeds, ranging from 2 Mpixel/s to up to 4 Mpixel/s.

\begin{figure}
    \centering
    \includegraphics[width=0.7\linewidth]{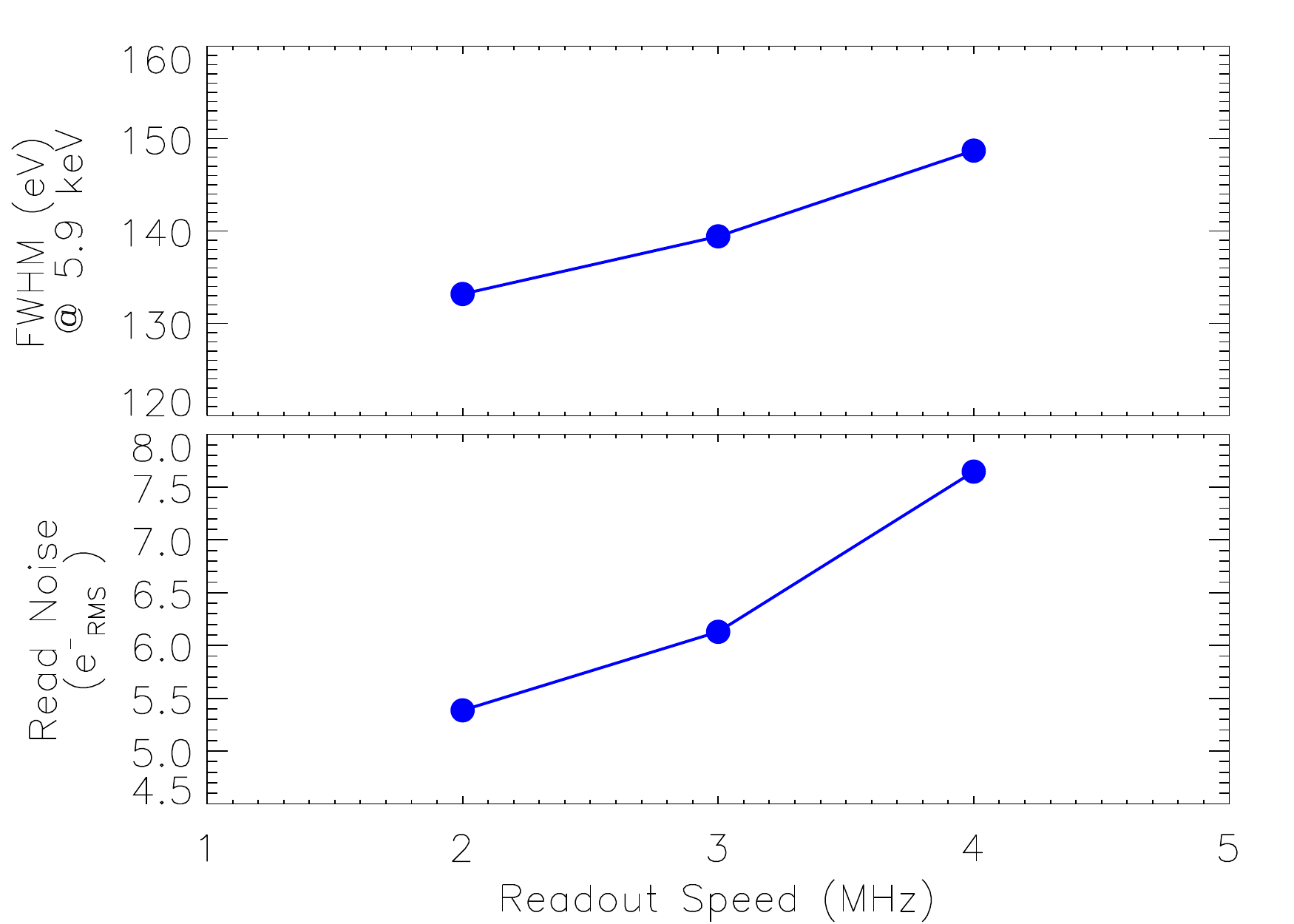}
    \caption{Spectral and noise performance of the device with readout speed.}
    \label{per_speed}
\end{figure}

Table \ref{tab1} summarizes the measurements at different readout speeds. 
\begin{table}[ht]
\caption{Summary of noise and spectral performance of the CCID85 devices} 
\label{tab1}
\begin{center}       
\begin{tabular}{|l|l|l|l|} 
\hline
\rule[-1ex]{0pt}{3.5ex}  Parameters & 2 MHz & 3 MHz & 4 MHz  \\
\hline\hline
\rule[-1ex]{0pt}{3.5ex}  Number of samples in CDS filtering &18 &10 &7    \\
\hline
\rule[-1ex]{0pt}{3.5ex}  Mean dark current ($e^{-}/s$) &50 &50 &50    \\
\hline
\rule[-1ex]{0pt}{3.5ex}  Read noise ($e^{-}_{RMS}$) &5.3&6.1&7.6  \\
\hline
\rule[-1ex]{0pt}{3.5ex}  Board noise ($e^{-}_{RMS}$) &1.7&2.0 & 3.2   \\
\hline
\rule[-1ex]{0pt}{3.5ex}  System gain ($ADU/e{^-}$) &0.84&0.83 &0.81    \\
\hline
\rule[-1ex]{0pt}{3.5ex}  Conversion gain ($\mu V/e^{-}$) &28&27.8&26.8    \\
\hline
\rule[-1ex]{0pt}{3.5ex}  FWHM (eV) @ 5.9 keV &133&139 &148    \\
\hline
\end{tabular}
\end{center}
\end{table} 
The spectral and noise performance of the device as a function of readout speed is shown in Fig. \ref{per_speed}. 
While the spectral resolutions for the grade 0 events are close to being Fano limited even at 4 MHz, read noise of the system can be further improved at higher readout speeds. 

While the output stages of the detectors need to be eventually made faster to support higher readout speeds, for the same detectors, we found that a slightly higher readout speed than 4 MHz can be achieved with better quality of the reset clock. 
We plan to improve the readout electronics module to read out these devices at speeds equal or exceeding 5 Mpixel/s with the use of a faster clock for the reset signal which will also lead to better noise performance of the detectors as discussed in Sec. \ref{speed}. For example, an LVDS module which is available in some models of the Archon controller, with temporal resolution of $\sim$1 ns, can be used for reset clock. An alternate approach might be to use a fast MOSFET clock driver to source buffer the reset clock. In parallel to these improvements, an ASIC based readout system development is ongoing that integrates the same readout module architecture as described in this paper to support parallel readout of larger arrays of X-ray CCDs \cite{herrmann20_mcrc}.


\subsection* {Acknowledgments}
We are deeply saddened by the passing of our colleague and co-author Barry Burke. His contribution has helped in the advancement of the X-ray CCD technology. We will miss him as a colleague, and above all as a kind human being.   

This work has been supported by APRA $\#$80NSSC19K0499 ``Development of Integrated Readout Electronics for Next Generation X-ray CCDs" and the SAT 80NSSC20K0401 ``Toward Fast, Low-Noise, Radiation-Tolerant
X-ray Imaging Arrays for Lynx: Raising Technology Readiness Further".




\begin{thebibliography}{10}

\bibitem{Lesser15_ccd}
M.~Lesser, ``A summary of charge-coupled devices for astronomy,'' {\em
  Publications of the Astronomical Society of the Pacific} {\bf 127}(957), 1097
   (2015).

\bibitem{gruner02_ccd}
S.~M. Gruner, M.~W. Tate, and E.~F. Eikenberry, ``Charge-coupled device area
  x-ray detectors,'' {\em Review of Scientific Instruments} {\bf 73}(8),
  2815--2842  (2002).

\bibitem{gaskin15_lynx}
J.~A. {Gaskin}, M.~C. {Weisskopf}, A.~{Vikhlinin}, {\em et~al.}, ``{The X-ray
  Surveyor Mission: a concept study},'' in {\em UV, X-Ray, and Gamma-Ray Space
  Instrumentation for Astronomy XIX},  {\em Society of Photo-Optical
  Instrumentation Engineers (SPIE) Conference Series} {\bf 9601}, 96010J
  (2015).

\bibitem{mushotzky2019_axis}
R.~F. Mushotzky, J.~Aird, A.~J. Barger, {\em et~al.}, ``{The Advanced {X}-ray
  Imaging Satellite},''  (2019).

\bibitem{lumb00_pileup_xmm}
D.~H. {Lumb}, ``{Simulations and Mitigation of Pile-Up in XMM CCD
  Instruments},'' {\em Experimental Astronomy} {\bf 10}, 439--456  (2000).

\bibitem{dan20_bkg}
D.~R. Wilkins, S.~W. Allen, E.~D. Miller, {\em et~al.}, ``{Identifying charged
  particle background events in X-ray imaging detectors with novel machine
  learning algorithms },'' in {\em Space Telescopes and Instrumentation 2020:
  Ultraviolet to Gamma Ray},  J.-W.~A. den Herder, S.~Nikzad, and K.~Nakazawa,
  Eds.,  {\bf 11444}, 308, International Society for Optics and Photonics, SPIE
   (2020).

\bibitem{falcone18_HXDI}
J.~A. Gaskin, A.~Dominguez, K.~Gelmis, {\em et~al.}, ``{The Lynx X-ray
  Observatory: concept study overview and status},'' in {\em Space Telescopes
  and Instrumentation 2018: Ultraviolet to Gamma Ray},  J.-W.~A. den Herder,
  S.~Nikzad, and K.~Nakazawa, Eds.,  {\bf 10699}, 120 -- 129, International
  Society for Optics and Photonics, SPIE  (2018).

\bibitem{bai08}
Y.~{Bai}, J.~{Bajaj}, J.~W. {Beletic}, {\em et~al.}, ``{Teledyne Imaging
  Sensors: Silicon CMOS imaging technologies for x-ray, UV, visible, and near
  infrared},''  {\bf 7021}, 702102  (2008).

\bibitem{hull17}
S.~V. {Hull}, A.~D. {Falcone}, D.~N. {Burrows}, {\em et~al.}, ``{Recent X-ray
  hybrid CMOS detector developments and measurements},''  {\bf 10397}, 1039704
  (2017).

\bibitem{chattopadhyay18_HCDoverview}
T.~Chattopadhyay, A.~D. Falcone, D.~N. Burrows, {\em et~al.}, ``X-ray hybrid
  cmos detectors: Recent development and characterization progress,'' in {\em
  Space Telescopes and Instrumentation 2018: Ultraviolet to Gamma Ray},  {\em
  Proc.SPIE} {\bf 10709}  (2018).

\bibitem{hull18_small_pixel}
S.~V. Hull, A.~D. Falcone, D.~N. Burrows, {\em et~al.}, ``Small pixel hybrid
  cmos detectors,'' in {\em Space Telescopes and Instrumentation 2018:
  Ultraviolet to Gamma Ray},  {\em Proc.SPIE} {\bf 10709}  (2018).

\bibitem{norbert16_wfi}
N.~Meidinger, J.~Eder, T.~Eraerds, {\em et~al.}, ``{The wide field imager
  instrument for Athena},'' in {\em Space Telescopes and Instrumentation 2016:
  Ultraviolet to Gamma Ray},  J.-W.~A. den Herder, T.~Takahashi, and M.~Bautz,
  Eds.,  {\bf 9905}, 662 -- 673, International Society for Optics and
  Photonics, SPIE  (2016).

\bibitem{SOI18}
T.~G. Tsuru, H.~Hayashi, K.~Tachibana, {\em et~al.}, ``{Kyoto's event-driven
  x-ray astronomy SOI pixel sensor for the FORCE mission},'' in {\em High
  Energy, Optical, and Infrared Detectors for Astronomy VIII},  A.~D. Holland
  and J.~Beletic, Eds.,  {\bf 10709}, 132 -- 142, International Society for
  Optics and Photonics, SPIE  (2018).

\bibitem{bautz18}
M.~Bautz, R.~Foster, B.~LaMarr, {\em et~al.}, ``{Toward fast low-noise
  low-power digital CCDs for Lynx and other high-energy astrophysics
  missions},'' in {\em Space Telescopes and Instrumentation 2018: Ultraviolet
  to Gamma Ray},  J.-W.~A. den Herder, S.~Nikzad, and K.~Nakazawa, Eds.,  {\bf
  10699}, 238 -- 248, International Society for Optics and Photonics, SPIE
  (2018).

\bibitem{bautz19}
M.~W. {Bautz}, B.~E. {Burke}, M.~{Cooper}, {\em et~al.}, ``{Toward fast,
  low-noise charge-coupled devices for Lynx},'' {\em Journal of Astronomical
  Telescopes, Instruments, and Systems} {\bf 5}, 021015  (2019).

\bibitem{bautz20}
M.~Bautz, B.~Burke, M.~Cooper, {\em et~al.}, ``{Progress toward fast,
  low-noise, low-power CCDs for Lynx and other high-energy astrophysics
  missions},'' in {\em Space Telescopes and Instrumentation 2020: Ultraviolet
  to Gamma Ray},  J.-W.~A. den Herder, S.~Nikzad, and K.~Nakazawa, Eds.,  {\bf
  11444}, 1318 -- 1323, International Society for Optics and Photonics, SPIE
  (2020).

\bibitem{Chattopadhyay20_spie}
T.~Chattopadhyay, S.~Herrmann, S.~Allen, {\em et~al.}, ``{Tiny-box: a tool for
  the versatile development and characterization of low noise fast x-ray
  imaging detectors},'' in {\em X-Ray, Optical, and Infrared Detectors for
  Astronomy IX},  A.~D. Holland and J.~Beletic, Eds.,  {\bf 11454}, 368 -- 385,
  International Society for Optics and Photonics, SPIE  (2020).

\bibitem{archon14}
G.~Bredthauer, ``{Archon: A modern controller for high performance astronomical
  CCDs},'' in {\em Ground-based and Airborne Instrumentation for Astronomy V},
  S.~K. Ramsay, I.~S. McLean, and H.~Takami, Eds.,  {\bf 9147}, 1730 -- 1740,
  International Society for Optics and Photonics, SPIE  (2014).

\bibitem{smith18_spie}
R.~Smith and S.~Kaye, ``{CCD speed-noise optimization at 1 MHz},'' in {\em High
  Energy, Optical, and Infrared Detectors for Astronomy VIII},  A.~D. Holland
  and J.~Beletic, Eds.,  {\bf 10709}, 229 -- 239, International Society for
  Optics and Photonics, SPIE  (2018).

\bibitem{herrmann20_mcrc}
S.~Herrmann, J.~Wong, T.~Chattopadhyay, {\em et~al.}, ``{MCRC V1: development
  of integrated readout electronics for next generation x-ray CCD detectors for
  future satellite observatories},'' in {\em X-Ray, Optical, and Infrared
  Detectors for Astronomy IX},  A.~D. Holland and J.~Beletic, Eds.,  {\bf
  11454}, 412 -- 418, International Society for Optics and Photonics, SPIE
  (2020).

\end{thebibliography}




\end{spacing}
\end{document}